\documentclass[%
 reprint,
 amsmath,amssymb,
 aps,
]{revtex4-2}

\usepackage{graphicx}
\usepackage{dcolumn}
\usepackage{bm}
\usepackage{svg}
\usepackage{hyperref}

\newcommand{\trace}{\text{Tr} }

\newcommand{\bea}{\begin{eqnarray}}
\newcommand{\eea}{\end {eqnarray}}
 
\newcommand{\Tr}{\textrm{Tr}\,}
\newcommand{\ads}{ \alpha_\textrm{DS}}

\begin{document}


\title{Testing dynamical stabilization of Complex Langevin simulations of QCD}

\author{Michael Westh Hansen}
 \email{michael.hansen@uni-graz.at}
\author{D\'enes Sexty}%
\affiliation{%
Institute of Physics, NAWI Graz, University of Graz,\\
Universitätsplatz 5, Graz, Austria }%



\date{\today}

\begin{abstract}
We study complex Langevin simulations of 
a toy model as well as QCD, supplemented
with a dynamical stabilization (DS) term, which was proposed 
to regularize the complexified 
process at lower temperatures. 
We compare the results to reweghting from zero
chemical potential to measure the bias that the 
inclusion of the stabilization term causes, depending 
on its strength.
At high temperatures the stabilization term
is not needed. At 
low temperatures (below deconfinement transition)
the DS term has a beneficial stabilizing effect, but 
too strong DS term causes phase quenching on the system. 
We observed that the bias of the 
dynamical stabilization can be to a good accuracy
removed by extrapolating to zero dynamical stabilization force
using a sigmoid fit.
\end{abstract}

\maketitle


\section{Introduction}

Investigations of the phase diagram of QCD, needed e.g.
to understand high density states of nuclear matter in neutron stars
are hindered by the sign-problem.
The sign problem is introduced by the fermionic degrees of 
freedom, which contribute a complex determinant in the measure
at nonzero chemical potential $\mu$.

 There have been many 
 methods proposed to circumvent the sign problem 
 of lattice QCD (and other theories).
The reweighting technique shifts the sign problem into 
the observable by choosing to simulate in a positive ensemble
\cite{Barbour:1997ej,Fodor:2001pe,Giordano:2020roi}.
The Taylor expansion uses simulations at $\mu=0$ to measure
derivatives of the partition function (which are charge densities 
and charge fluctuations) to extrapolate to finite $\mu$
\cite{Allton:2002zi,Gavai:2008zr,Borsanyi:2011sw,Bazavov:2017dus}.
In the analytical continuation method, one uses imaginary 
chemical potentials for simulations, as the theory is again positive there,
and in effect extrapolates results from $ \mu^2 \le 0 $ to $ \mu^2>0$
\cite{deForcrand:2002hgr,DElia:2002tig,Cea:2014xva,Bellwied:2015rza}.

The general behavior of the above mentioned extrapolation methods is 
that as long as we are interested in small chemical potentials,
they give consistent and reliable results, and their range of 
applicability depends on the temperature $T$.
Around $ \mu/T \simeq 1$
(corresponding to $ \mu_B /T \simeq 3$ with the baryon
chemical potential $\mu_B$), results start to deteriorate, the 
different methods give inconsistent results.
Further methods used with varying amount of success are the density of states 
methods \cite{Fodor:2007vv,Endrodi:2018zda},
canonical ensembles \cite{Alexandru:2005ix,Kratochvila:2005mk}
and dual formulations \cite{Gattringer:2014nxa}.
For more details and further references, see the reviews \cite{Gattringer:2016kco,Guenther:2020jwe}.

The next group of methods exploits the complex analytical properties 
of the theory: the complex Langevin method, the Lefschetz-thimble and related methods.
The Complex Langevin method \cite{Klauder:1983nn,Parisi:1984cs} 
is a straightforward generalization of the real-Langevin method (a.k.a 
stochastic quantisation) for complex actions, and it does not 
rely on the interpretation of the measure $ e^{-S} $ as a probability, 
therefore it circumvents the sign problem. The complex action
in turn leads to the complexification of the field manifold, and 
thus a stochastic process on this enlarged space, for which an analytic continuation
is used to recover the original theory.
One drawback of the complex Langevin method is that in some cases 
it gives incorrect results. The theoretical understanding of the 
failure modes have improved 
significantly
\cite{Aarts:2009uq,Aarts:2009dg,Aarts:2011ax,Aarts:2012ft,Salcedo:2018fvt,Cai:2020tgd,Seiler:2023kes} and diagnostic tools such as 
boundary terms \cite{Scherzer:2018hid,Scherzer:2019lrh,Seiler:2020mkh} were introduced as well as other 
diagnostic observabes \cite{Nagata:2016vkn,Nagata:2018net}.
Recently it was 
shown that the wrong convergence in some cases can be alleviated 
using the kernel freedom of the Complex Langevin method
\cite{Alvestad:2022abf,Boguslavski:2022dee,Lampl:2023xpb,Alvestad:2023jgl},
which itself uses modern machine learning inspired techniques.


In naive complex Langevin simulations of gauge theories one observes 
an uncontrollable growth of fluctuations as the fields
explore the non-compact complexified manifold of the gauge degrees of freedom.
This is negated by gauge-cooling \cite{Seiler:2012wz,Aarts:2013uxa,Nagata:2015uga,Cai:2019vmt,Dong:2020mtk}, which uses the 
gauge degree of freedom to keep the fields close to the real manifold.
This in turn makes CLE simulations of gauge theories possible,
and it was used to map the whole phase diagram of HDQCD \cite{Aarts:2016qrv}.
It was shown that with gauge-cooling one can simulate 
full QCD at high chemical potentials \cite{Sexty:2013ica,Aarts:2014bwa,Fodor:2015doa,Nagata:2016mmh,Nagata:2018mkb,Kogut:2019qmi,Attanasio:2020spv,Yokota:2023osv}. 
At high temperatures the equation of state was also measured 
at large chemical potentials \cite{Sexty:2019vqx,Attanasio:2022mjd}, although not yet at physical quark masses.

Dynamical stabilization was proposed to 
help complex Langevin simulations \cite{Attanasio:2018rtq} (see also \cite{Cai:2021hjq}),
keeping their fluctuations in imaginary directions small.
It is an ad hoc addition to the drift terms 
of the theory, which present a force pushing the fields towards the 
real manifold. It can also be viewed as a soft cut off in imaginary directions which 
limits the deviation of the fields from the real manifold.
It was used in context of the Heavy Dense QCD as well as for full QCD 
\cite{Attanasio:2022mjd}.
In this paper we study the effects of dynamical stabilization on a toy model,
and in QCD simulations. We quantify its effect on the results of the CLE
simulations as a function of the strength of the stabilization force, and demonstrate
that its functional dependence is well described with by 
a sigmoid fit and thus extrapolation to zero dynamical stabilization
allows to remove most of the effects of the stabilization on the results.

In section 2 we give a short introduction to the Complex Langevin method and the boundary terms 
as well as the dynamical stabilization, and describe the
systems we investigate.
In section 3 we describe our numerical results, and finally in section 4 we conclude.

\section{Complex Langevin  eq. and numerical setup}
In this paper we investigate models with non-zero chemical potential, which 
means the action $S$ as well as the measure $ e^{-S} $ is in general complex. 
This invalidates naive importance sampling simulations, such as the Metropolis-Hastings algorithm. This problem is known 
as the sign problem, as the accept/reject step
is rendered undefined due to negative- (and in general complex-) valued probabilities.

It is possible to go around the sign problem in some cases using the reweighting method, which
pushes the non-positive part of the measure into the observable.
This works well if the sign problem is mild (the chemical potential is small enough).
For a severe sign problem reweighting becomes unusable due to a signal-to-noise ratio
problem. Due to the overlap problem observables will be dominated by very few configurations,
which means that the distribution of the observables is far from a Gaussian 
and thus naively calculated errorbars are also untrustworthy.

For larger values of the chemical potential one has to investigate different methods, and here we will look into the Complex Langevin approach. 
This is based on stochastic quantisation, which we 
briefly introduce below. In this approach we form a stochastic process,
which equilibrates to the measure we are interested in, namely 
$ \rho(x) \propto e^{-S(x)}$ for some real action $S(x)$.
This can be achieved using the stochastic differential equation
called Langevin equation which is set up for the variables of our action.
Written for a single degree of freedom it is given by
\bea \label{langevineq}
  dx(\tau) = K(x) d \tau + dw_\tau,
\eea
where $\tau$ is the Langevin time (similar to MC-time in usual importance sampling 
simulations),
the drift term $K[x]$ is calculated from the action $K[x]=-\partial_x S$, and 
$dw_\tau$ is the increment of a Wiener-process with variance $\langle dw(t)^2 \rangle = 2 dt $.
This stochastic process implies the Fokker-Planck equation for 
the probability density $ P(x,\tau)$: 
\begin{equation}
    \frac{\partial}{\partial t} P(x,\tau) = \frac{\partial}{\partial x} \left[\left( \frac{\partial}{\partial x} - K(x) \right) P(x,\tau) \right].
\end{equation}
It is easy to see that the stationary measure $P(x)=e^{-S(x)}$
satisfies this equation, and one can show that except for 
certain special cases this is a unique stationary state and the 
process equilibrates to it for real actions \cite{Damgaard:1987rr}.
On the right hand side of the Fokker-Planck equation 
the operator acting on $P(x,\tau)$ is also known as the Langevin operator. 

Thus the Langevin equation then forms our stochastic process, which can be discretized 
using e.g. the Euler-Maruyama scheme, 
\begin{equation}
    x(\tau + \epsilon) = x(\tau) + \epsilon K[x(\tau)] + \sqrt{\epsilon} \eta(\tau)
\end{equation}
where $\eta$ is a Gaussian noise with zero average and variance of two. 

Even for a complex action (equivalently complex measure), one can use the 
same Langevin equation (\ref{langevineq}) to 
define a stochastic process. The field gets complexified, and we use the analytically 
continued drift terms to define $K(x)$ on the complexified manifold \cite{Klauder:1983nn,Parisi:1984cs}.

In the case of gauge-theories our variables will be elements of a Lie-group, 
corresponding to parallel transporters along the links of a cubic space-time 
lattice.
We will be interested in SU($N$) gauge theories below, and in particular in $N=3$ describing
the strong interaction sector of the Standard model.
The derivative in the drift term in this case is the left derivative, with respect to the Lie-group's generators, such that
\begin{equation}
    D_a f (U) = \left. { \partial\over \partial \alpha } f( e^{i \alpha \lambda_a} U) \right|_{\alpha=0}  
    \label{SU_der}
\end{equation}
where $\lambda_a$ is the a'th Gell-Mann matrix for $N=3$. 
Defining our gauge theory on a cubic space-time lattice, the drift term for the link variable $U_\nu(x) $ at site $x$ in direction 
$\nu$ is given by
\bea
K^a_{\nu x} = -D_{\nu x a} S[U].
\eea
The discretised Langevin equation becomes 
\begin{equation} \label{gaugelangevinupdate}
    U_\nu^{\tau+\epsilon}(x) = \exp \left[i\lambda_a \left( \epsilon K^a_{\nu x} [U^{\tau}] + \sqrt{\epsilon} \eta^a_{\nu x}[U^{\tau}] \right)\right] U_\nu^{\tau}(x).
\end{equation}
The use of the exponential update factor ensures 
that the determinant of the $U$ matrices remain unity, however
it is worth noting that if the action is complex, then the drift term $K^a_{\nu x}[U]$ is also complex. This means the SU($N$)-element
link variables drift into SL($N,\mathbb{C}$), 
as the unitarity is no longer preserved by the update.

\subsection{Boundary terms}

One of the failure modes of the Complex Langevin simulations 
is caused by the complexified distributions haveing 
a long tail in imaginary directions.
This than invalidates the formal proof of correctness \cite{Aarts:2009uq,Aarts:2011ax}.
The discrepancy of the correct results and Complex Langevin results 
can be understood in terms of boundary terms as follows 
\cite{Scherzer:2018hid,Scherzer:2019lrh,Seiler:2020mkh}:
We define an interpolation function, which for an action
depending on a single variable is written as
\begin{equation}
    F_\mathcal{O}(t,\tau) = \int P(x,y,t-\tau) \exp(\tau L_c) \mathcal{O}(x,y) dxdy    
\end{equation}
where $P(x,y,t)>0$ is the distribution of the process on the 
complexified manifold at Langevin time $t$, $O(x,y) = O(x+iy)$ 
is some holomorphic observable, and $L_c$ is the 
complex Fokker-Planck operator, which (for theories with 
multiple degrees of freedom) is written as 
\bea \label{lcop}
L_c= \sum_i ( \partial_i + K_i ) \partial_i, 
\eea
where $K_i$ is the drift term of the $i$-th field in the action and $ \partial_i$ is 
derivation with respect to the $i$-th field.
The combination $ \exp(\tau L_c) \mathcal{O}(x,y) $ thus defines a $\tau$ 
dependent observable.
Trivally, $ F(t,\tau=0) $ gives the expectation value of observable $\mathcal{O}(x,y)$
in the complex Langevin simulation at Langevin time $t$.
One can show that if one takes a large $ \tau=t$, one has 
with $ \rho = \exp(-S) $
\begin{equation}
    F_\mathcal{O}(t,t) = \int P(x,y,0) \exp(t L_c) \mathcal{O}(x,y) dxdy    = 
    \langle \mathcal{O} \rangle_{\rho},
\end{equation}
i.e. the correct result, assuming the $L_c$ operator has 
a non-degenerate zero eigenvalue and 
no eigenvalues with positive real part
\cite{Aarts:2009uq,Aarts:2011ax}. This spectral reqiurement 
is often fulfilled, however there are some known exceptions
\cite{Seiler:2023kes}.

We can thus take $ F(t,0)-F(t,t) $ as the difference of 
the CLE result and the correct result.
We define the $n$-th boundary term of the 
observable $\mathcal{O}$ as 
\bea
 B_n ( \mathcal{O}) = \partial_\tau^n F(t,\tau=0).
\eea 
In practice we mostly check the first boundary term 
of an observable, and we can take 
that value as an indicator of the magnitude of the errors
of Complex Langevin results. To calculate
this derivative we need to regularize the observable by 
an imaginary cutoff,
\begin{equation}
    B_1(\mathcal{O},C) = \int dx \int_{-C}^{C} dy P(x,y,0) (L_c \mathcal{O}(x,y)) 
\end{equation}
and measure the observable as an extrapolation to $ C \rightarrow \infty $.
(The boundary term  might be undefined or very noisy if 
one uses directly $ C = \infty$.)
This procedure is straightforwardly generalizable 
to lattice systems, where the generalization 
involves calculating the boundary term observables 
$L_c^n \mathcal{O}$ using (\ref{lcop}), and 
choosing a cutoff such that a finite $C$ defines a 
compact manifold incorporating the original real manifold.
For simulations using link variables
on SU(3), the complexified manifold
becomes SL(3,$\mathbb{C}$), and one can 
use the unitarity norm for the cutoff:
\bea
N_U=  \max_{x,\nu} \Tr \left[ ( U_{x\nu } U_{x \nu } ^{+} -1  )^2 \right] ,
\eea
such that the boundary term is given by
\begin{equation}
    B_n(\mathcal{O},C) =  \langle \theta( C-U_N ) L_c^n \mathcal{O} \rangle.    
\end{equation}


\subsection{Dynamical stabilization}
In some cases the 
large fluctuations of the fields in imaginary directions have been identified 
to cause wrong convergence of the Complex Langevin method.
Here we will investigate dynamical stabilization \cite{Attanasio:2018rtq}, which is 
essentially a soft cutoff on the distance of the link variables 
from the original SU(3) manifold. This is achieved by  
introducing a purely imaginary additional force in the drift 
term that pushes the variables towards SU(3).
This new term in the drift is not a derivative of an action,
 and moreover it is given by a non-holomorphic 
 function of the complexified fields of the theory.
This implies that the argument for correctness of the Complex Langevin method no longer applies,
indeed our aim is to quantify the bias that this addition imparts on our results.
The proposed force is given in \cite{Attanasio:2018rtq} as,
\begin{equation} \label{mixingdynstab}
    K_{x\nu}^{a} \rightarrow K_{x\nu}^{a} - i \alpha_{DS} b^a_x 
       \left( b_x^c b_x^c \right)^3
\end{equation}
with
\begin{equation} \label{b_x^a}
    b_x[U] = \text{Tr}\left(\lambda_a \sum_\nu U_{x\nu}^\dagger U_{x\nu} \right) 
\end{equation}
such that the force is zero if the gauge links are  SU(3)-elements, and preserves the gauge symmetry 
with respect to SU(3) gauge transformations. Notice that in $(b_x^c b_x^c)$, the sum over the Gell-Mann matrices can be simplified using the Fierz completeness relation of the generators for SU(N), giving
\begin{equation}
    b^c_x[U] b^c_x[U]= \text{Tr}\left(\sum_\nu U_{x\nu}^\dagger U_{x\nu} \right)^2 - \frac{1}{3} \text{Tr}^2\left(\sum_\nu U_{x\nu}^\dagger U_{x\nu} \right).
\end{equation}

The parameter $\alpha_{DS}$ controls the strength of the force and we will show results for multiple orders of magnitude of $\ads$, to get insight in this parameter's effect on the simulation. In addition we also investigate a slightly different dynamical stabilization force, which doesn't sum over the directions, to keep the force on link variables in different directions independent. 
\begin{equation} \label{nonmixingdynstab}
    K_{x\nu}^{a} \rightarrow K_{x\nu}^{a} - i \alpha_{DS} b^a_{x\nu}  \left( b_{x\nu}^c b_{x\nu}^c \right)^3,
\end{equation}
using
\begin{equation}
    b_{x\nu}[U] = \text{Tr}\left(\lambda_a U_{x\nu}^\dagger U_{x\nu} \right). 
\end{equation}
Also in this case one can use the Fierz complectness relations to simply the terms in the bracket:
\begin{equation}
    b^c_{x\nu}[U] b^c_{x\nu}[U]= \text{Tr}\left(U_{x\nu}^\dagger U_{x\nu} \right)^2 - \frac{1}{3} \text{Tr}^2\left(U_{x\nu}^\dagger U_{x\nu} \right).
\end{equation}

\subsection{One-link toymodel}

Before simulating full QCD, we will investigate a simple one-plaquette model
called Polyakov chain \cite{Seiler:2012wz}, here with only one link 
variable $U$
representing the Polyakov loop, with the action
\begin{equation} \label{toymodel}
    -S = \beta_1 \text{Tr} U + \beta_2 \text{Tr} U^{-1} 
\end{equation}
and 
\begin{equation}
    \beta_1 = \beta + \kappa \exp(\mu), \quad \beta_2 = \beta + \kappa \exp(-\mu)  
\end{equation}
Here $\beta$ and $\kappa$ are coupling constants, and $\mu$ is the chemical potential. Furthermore the drift term can be calculated to be
\begin{equation}
    K_a = i \beta_1 \text{Tr} \lambda_a U - i \beta_2 \text{Tr} \lambda_a U^{-1},
\end{equation}
using the left derivative described in equation \ref{SU_der}. 
We then add the dynamical stabilization to the drift, 
\bea \label{toymodeldrift}
    M_{a} & =  & K_{a} + S_a \nonumber \\
    S_a&=&i \alpha_{DS} d^3[U] \text{Tr}\left(\lambda_a U^\dagger U \right) 
\eea
with 
\begin{equation}
    d[U] = \text{Tr}\left(U^\dagger U \right)^2 - \frac{1}{3} \text{Tr}^2\left(U^\dagger U \right),
\end{equation}
and thus  we use $M_a$ instead of $K_a$ in the Langevin update equation 
(\ref{gaugelangevinupdate}).

For this model we have then found values for $\beta$, $\kappa$ and $\mu$, such that the model gives incorrect convergence without the use of dynamical stabilization. The observables 
considered here are the trace of the Polyakov loop $P=\Tr U$, and the trace of the inverse Polyakov loop $P '=\Tr U^{-1}$,
 which is then compared to the exact value. The exact 
 values are found by numerically integrating 
over conjugacy classes of the SU(3) group \cite{Aarts:2008rr},
such that the Polyakov loop is represented 
by a diagonal matrix $ \textrm{diag}( e^{i\phi_1},e^{i\phi_2},e^{-i(\phi_1 + \phi_2}))$ using the reduced Haar-measure of the conjugacy classes
\begin{eqnarray}
    dU &= &\sin^2\left(\frac{1}{2} \phi_1 + \phi_2\right) \sin^2\left(\phi_1 + \frac{1}{2} \phi_2\right) \cdot \\ \nonumber
   && \sin^2\left( \frac{\phi_1 - \phi_2}{2}\right) \frac{  d\phi_1 d\phi_2 }{N}
\end{eqnarray}
where $N$ is a normalizing factor, and $-\pi < \phi_1, \phi_2 \le \pi$ are the two free variables parametrizing the conjugacy classes.
This measure can be used in numerical integrations of  
the expectation values for $P$ and $P'$ using e.g. the trapezoidal rule. 

\subsection{QCD}

The full-QCD simulations use the Wilson plaquette action for the gauge degrees of the freedom and $N_F=4$ species of staggered fermions 
(if not noted otherwise).
The contribution of the gauge degrees of freedom to 
the drift terms is calculated straightforwardly from the action. (It is worth noting that writeing the 
plaquette action one uses $U^{-1}$ instead of $U^\dagger$, to preserve the holomorphicity of the action. 
While for $SU(N)$ this would be equivalent, extending the process to SL($N,\mathbb{C}$), it 
becomes a crucial distinction.)
We use gauge-cooling to avoid the fast blowup of the unitarity norm, enabled by the complexified gauge degrees of freedom \cite{Seiler:2012wz,Aarts:2013uxa}.
The contribution of the fermions to the $a$-th color component of the drift term of the link at site $x$, pointing in direction $\nu$ is written as 
\bea 
K^F_{x\nu a} = {N_F \over 4 }\trace ( M^{-1}  D_{x\nu a} M ).
\eea
with the Dirac matrix $M$, which for unimproved staggered fermions
reads as 
\bea
M_{xy} = m \delta_{xy} + \sum_\nu \eta_\nu(x) \left[ e^{\delta_{\nu 4}\mu} U_\nu(x) \delta_{x+\hat \nu,y}
\right. \nonumber \\  \left.
 - e^{-\delta_{\nu 4} \mu} U_\nu^{-1}(y) \delta_{x-\hat\nu,y} \right]
\eea
with the staggered phases $\eta_\nu(x)$.
The trace in the drift term is calculated using the noisy estimator \cite{Sexty:2013ica}, to avoid having 
to calculate the inverse of the Dirac matrix $M^{-1}$.
For small lattices the calculation of the inverse of the Dirac matrix is affordable (using 
e.g. the LU decomposition), so
in some cases we calculate the exact fermionic drift terms using the exact $M^{-1}$.
While more expensive, exact drift terms are beneficial as they allow for simulations at lower inverse gauge coupling ($\beta$) values,
because the lack of the noise in the drift terms results in more stable simulations.

\section{Results}

\subsection{One-link toymodel results}

In this section we first show numerical results for the toy model defined in eq. (\ref{toymodel}).
We used an adaptive Langevin update with a maximum stepsize of $ \Delta \tau=10^{-4}$. 
We typically used $ \tau= 100$ Langevin time for thermalization, and used further 
$ \tau=10^5$ for measuring the averages. To calculate the errorbars 
we used the Bootstrap procedure after binning the data to get rid of the autocorrelations
of the process. We chose $\beta=0.1$, $\kappa=0.25$ and $\mu=1.0$, which is a point where the Complex 
Langevin method gives slightly incorrect results. At larger $\beta$ and
lower $\kappa$ 
it gives correct results and thus dynamical stabilization is not needed.

\begin{figure}
    \centering
    \includegraphics[width=0.9\columnwidth]{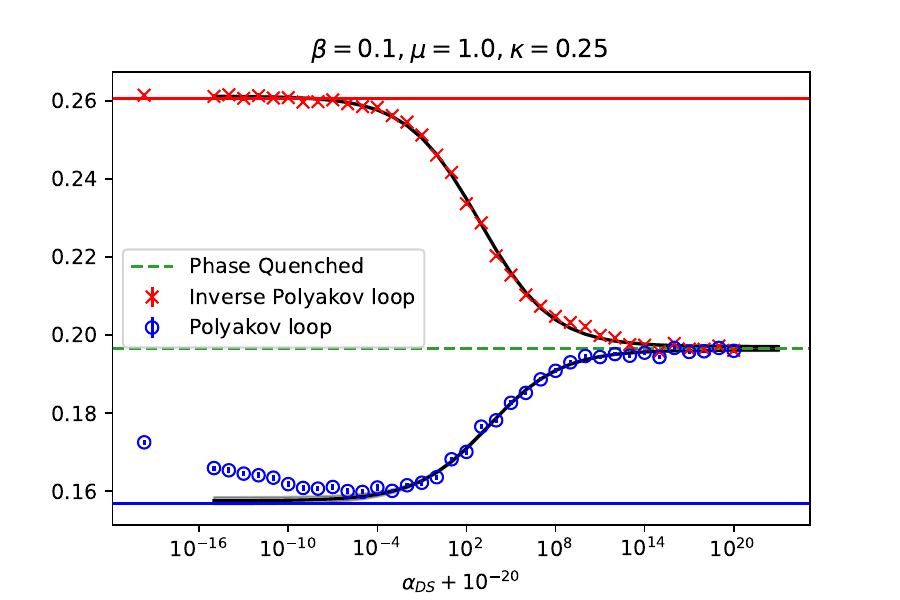}
    \includegraphics[width=0.9\columnwidth]{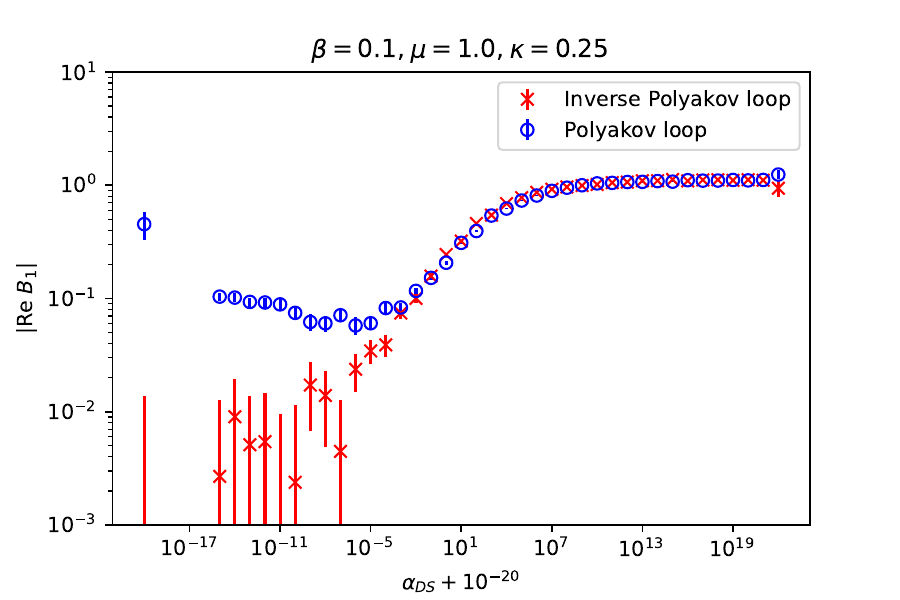}
    \caption{Above, the Polyakov loop $P$ and the inverse 
    Polyakov loop $P'$ are shown in the toy model (\ref{toymodel})
    at $\beta=0.1$,
    calculated using dynamically stabilized CLE, as a function of 
    the $ \alpha_\textrm{DS}$ parameter.
    The exact 
    results calculated by numerical integration (red/blue lines) and 
    the phase quenched exact result (green line) is also shown, as well as a fitted 
    sigmoid curve, 
    defined in (\ref{fitfunction}) (dark gray line with error bands).
    Below, the absolute values of the boundary terms are shown 
    for the Polyakov loop and the inverse Polyakov loop variable as a function
    of $\ads$.
   }
    \label{fig:testmod1}
\end{figure}

We first test the dependence of the results on the control parameter of 
the dynamical stabilization $ \alpha_\textrm{DS}$ in Fig.~\ref{fig:testmod1}.
We show the numerically integrated exact results as well 
as the phase quenched exact value (PQ), which is calculated by taking 
the absolute value of the measure in the expectation value:
\bea
 \langle P \rangle_\textrm{PQ} &=& {1\over Z_\textrm{PQ}} \int P e^{-\textrm{Re} S(U)} dU, \\
 \nonumber
 Z_\textrm{PQ} &=& \int e^{-\textrm{Re} S(U) }dU.
\eea
The behavior of the process changes qualitatively if $\ads$ is changed 
over a large range of magnitudes, therefore we plot results on a logarithmic $\ads$ scale.
First note that at $\ads=0$ the complex Langevin results are incorrect,
we have checked that this is correctly signalled also by nonzero 
boundary terms, as visible on the second plot of Fig.~\ref{fig:testmod1}.
At large $\ads$ the link variable has a strong force pushing 
it back towards the SU(3) manifold. it is expected that in the limit of very large $\ads$,
the link variable will be confined to the SU(3) manifold, and its dynamics 
will be given by that of phase quenched theory, as the real parts of the drift 
term are calculated from the action, and the imaginary parts are rendered mute by the 
dynamical stabilization term. In the phasequenched theory the real part of the Polyakov loop 
and its inverse are equal, as the link variable is unitary. This
behavior is nicely observed in the numerical experiments,
as visible in Fig.~\ref{fig:testmod1}.
Also noteworthy is the fact that even though the CLE results are incorrect at $ \ads=0$,
introducing a moderate value for $\ads$ actually drives the results 
closer to the exact value, before finally they approach the phasequenched value.
The $\ads$ dependence of the results is well described 
by a sigmoid dependence if we exclude the small $\ads$ region. We have fitted the data with a 
four parameter fit using the function
\bea \label{fitfunction}
f(\ads) = A+ {B - A\over 1 + C \ads^D }.
\eea
For the fit we only used simulation 
values at $ \ads \ge 10^{-5} $ (Not knowing 
the exact results, one could justify this by the 
non-monotonic behavior as a function of $\ads$
or the increase of the boundary terms,
as well as the 
increase in the  
$R$ ratio (defined below).)
The quality of the 
fit can be judged from $\chi^2/n_\textrm{DOF} \approx 1-3 $ for this fitting range.  
As observed in Fig.~\ref{fig:testmod1},
the fit function extrapolated to $\ads=0$ gives
the correct results with good accuracy, as also visible in 
Table~\ref{tab:toymodel}.
\begin{table}
\begin{tabular}{|c|c|c|}
\hline
     & $\langle U \rangle $ &  $\langle U^{-1} \rangle $ \\
\hline
exact &  0.15744528  & 0.26051165  \\ 
extrapolated &  0.15756(94)  &   0.26113(25)  \\ 
CLE $ \ads =0$ & 0.17254(61) & 0.26139(62) \\   
\hline
\end{tabular}
\caption{Comparison of the exact values, 
CLE results extrapolated to $\ads=0$, and 
CLE results of simulations at $\ads=0$.
}
\label{tab:toymodel}
\end{table}
Notably, for the parameter $D$ (the exponent of $\ads$ in
the sigmoid function) we get a value 
close to 0.25.

\begin{figure}
    \centering 
    \includegraphics[width=0.9\columnwidth]{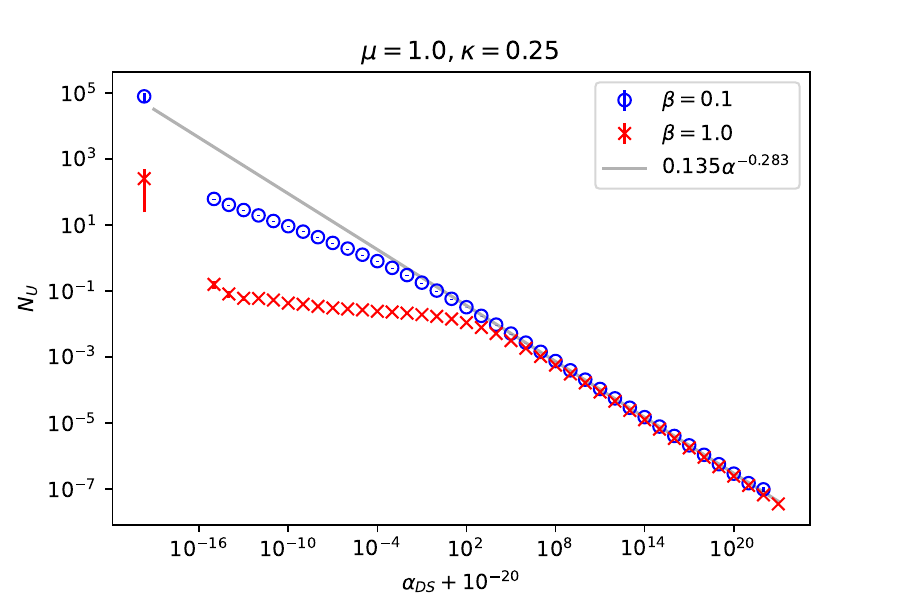}
    \includegraphics[width=0.9\columnwidth]{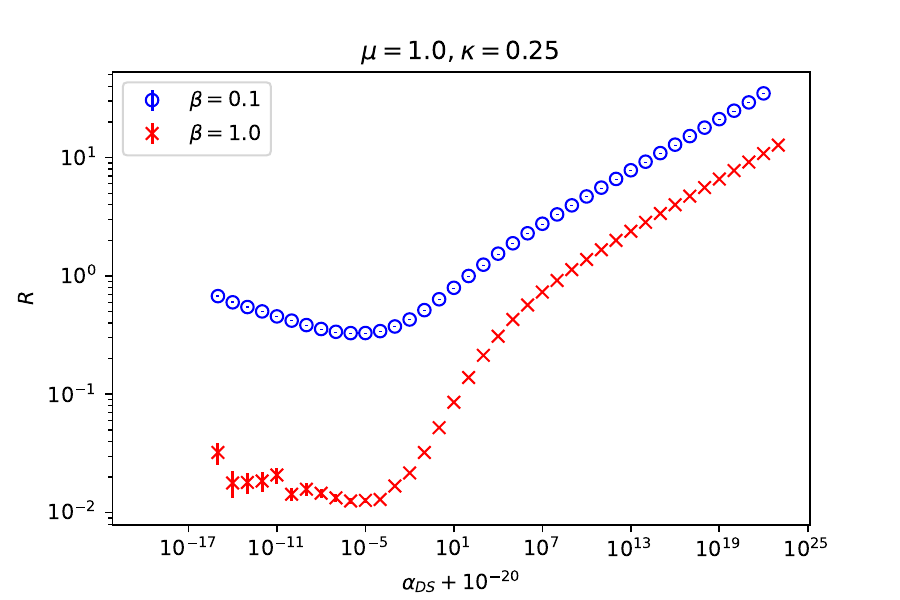}
    \caption{The unitarity norm as well as a 
    the ratio of the norm of the drift terms coming from the action vs the norm of the 
    dynamical stabilization force (defined in (\ref{Rdefinition})) as a 
    function of $\ads$, for the toy model. 
    }
    \label{fig:testmod2}
\end{figure}

\begin{figure}
    \centering
    \includegraphics[width=0.9\columnwidth]{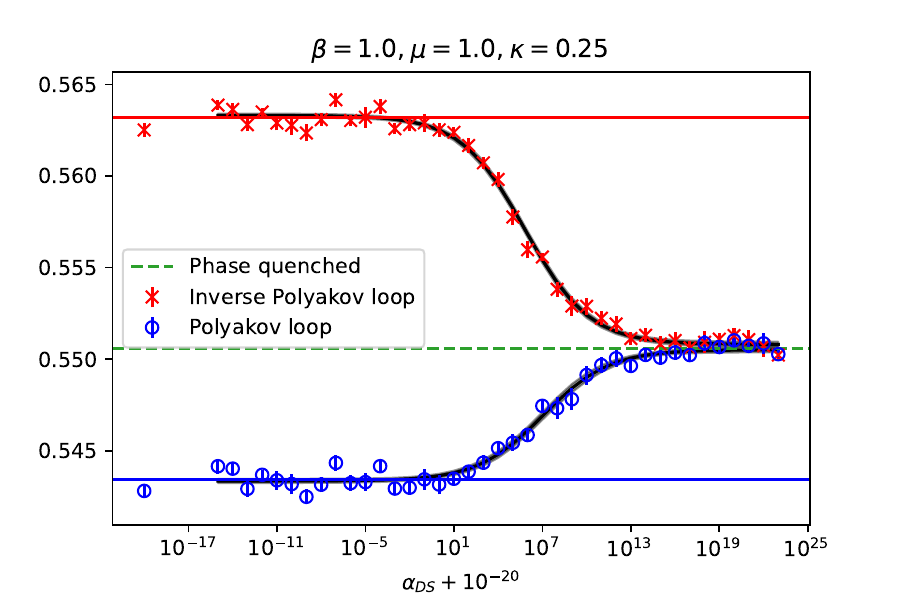}
    \includegraphics[width=0.9\columnwidth]{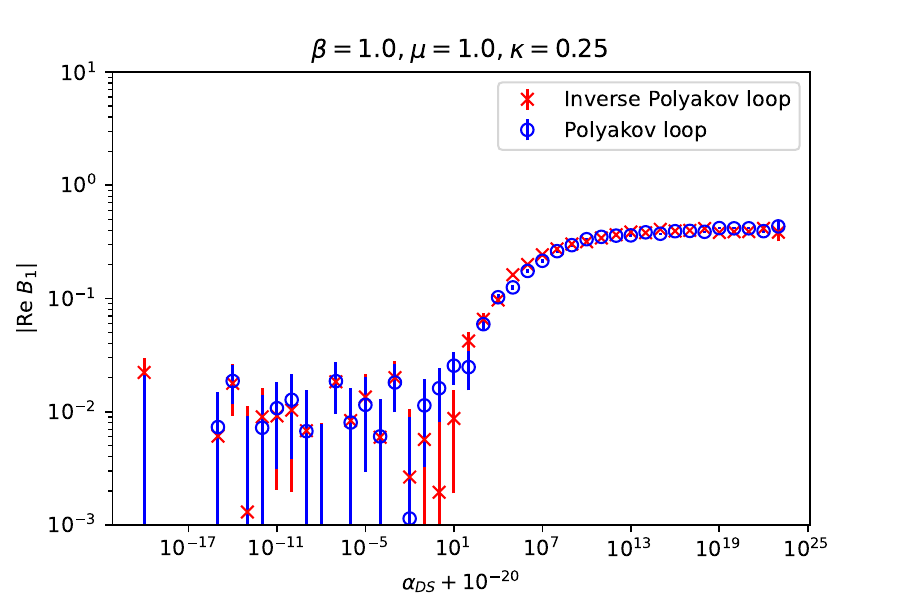}
    \caption{Above, the Polyakov loop $P$ and the inverse 
    Polyakov loop $P'$ are shown in the toy model (\ref{toymodel})
    at $\beta=1.0$,
    calculated using dynamically stabilized CLE, as a function of 
    the $ \alpha_\textrm{DS}$ parameter.
    The exact 
    results and the phase quenched result, are also shown (full lines).
    Below, the absolute value of the boundary terms are shown 
    for the Polyakov loop and the inverse Polyakov loop variable as a function
    of $\ads$.
   }
    \label{fig:testmodhighb}
\end{figure}

In Fig.~\ref{fig:testmod2} we show the unitarity norm 
\bea
N_U=\Tr \left[ ( U U ^{\dagger} -1  )^2 \right]
\eea
as well as the ratio of norms of the dynamical stabilization term
versus the drift term coming from the action
\bea \label{Rdefinition}
R = {\sqrt { \langle  \sum_a |S_a |^2  \rangle }
\over \sqrt {\langle \sum_a |K_a |^2 \rangle } }.
\eea 
This gives information on how much the stabilization 
changes the original drift terms of the theory.
The magnitude of the norm of $K_a$ is approximately independent of $\ads$ so 
the ratio roughly corresponds to the norm 
of $S_a$.
We observe that this ratio has a minimum in the investigated $\ads$
range, as for small $ \ads $ the fields deviate more from SU(3), and thus 
the attractive force towards SU(3) tends to be larger.
For large $\ads$ the force increases again as it is 
multiplied by a factor $\ads$.
Note that the minimum of this ratio roughly coincides with 
the $\ads$ value which corresponds to minimal boundary terms,
as visible in Fig.~\ref{fig:testmod1}.
We also carried out simulations at $\beta=1.0$, where the unstabilized CLE
simulations already gave a correct result. A small stabilization
force does not spoil this, but a too large $\ads$ drives 
the results to the phasequenced values,
as visible in Fig.~\ref{fig:testmodhighb}.
At $\beta=1.0$ both the direct simulation and the extrapolation to $ \ads=0$
from simulations above $ \ads=10^{-5}$ (where $R$ has a minimum) give
results which are consistent with exact results within statistical errors.
For simulations at $\beta=1.0$ we got $ \chi^2 / n_\textrm{DOF} \approx 1-1.3$, showing 
that the $\ads$ dependence is well described by the ansatz.
At lower $\ads$ values for $\beta=1.0$, the unitarity norm levels 
off at a certain value, as the distribution is 
constrained by the complexified stochastic process, and not by the stabilizing force,
as observed in Fig.~\ref{fig:testmod2}.
At high $\ads$, the decay of the unitarity norm 
follows a power law behavior with the exponent $ \approx -0.28$.

The boundary terms also confirm correct results at small $\ads$ values,
however at large $\ads$ they signal the incorrect results as 
the large dynamical stabilization force spoils the correspondance of 
the Complex Langevin process and the original complex measure.

The results in this section are qualitatively similar to our results in QCD at high 
temperature, to be discussed in the next section.

\subsection{QCD results}

Next we turn to results of simulations of full QCD. 
We used simulations on $ 8^3 \times 4 $ lattices (unless
otherwise noted)
using the plaquette action for the gauge fields and $ N_F=4$ flavors of 
unimproved staggered 
fermions using $m=0.02$ (in lattice units). 
To test different temperatures 
and thus different phases 
of the theory we change the $ \beta$ parameter of the gauge action.
After performing a $\beta$ scan to explore the qualitative 
behavior of the dynamical stabilization force,
we used $\beta=4.9$ for simulations in the deconfinement region
and $\beta=5.2$ for simulations in the confinement region.
(The critical inverse coupling on the $8^3 \times 4$ lattice
using the parameters mentioned above and $\mu=0$ is roughly $\beta_c\approx 5.0$.)

We use a nonzero chemical potential $\mu$ to induce a sign-problem.
For this study we use a relaticely small quark chemical potential $\mu=0.1$ (in lattice units), that 
is in a region where one can still calculate reweighted results 
with a reasonable effort with relatively small errors.
For the reweighting we used about 22000 configurations for $\beta=4.9$ and 6000 configurations for $\beta=5.2$  
which we collected using a HMC procedure at $\mu=0$ and 
the respective gauge couplings $\beta=4.9$ and $\beta=5.2$.

\begin{figure}
    \centering
    \includegraphics[width=0.9\columnwidth]{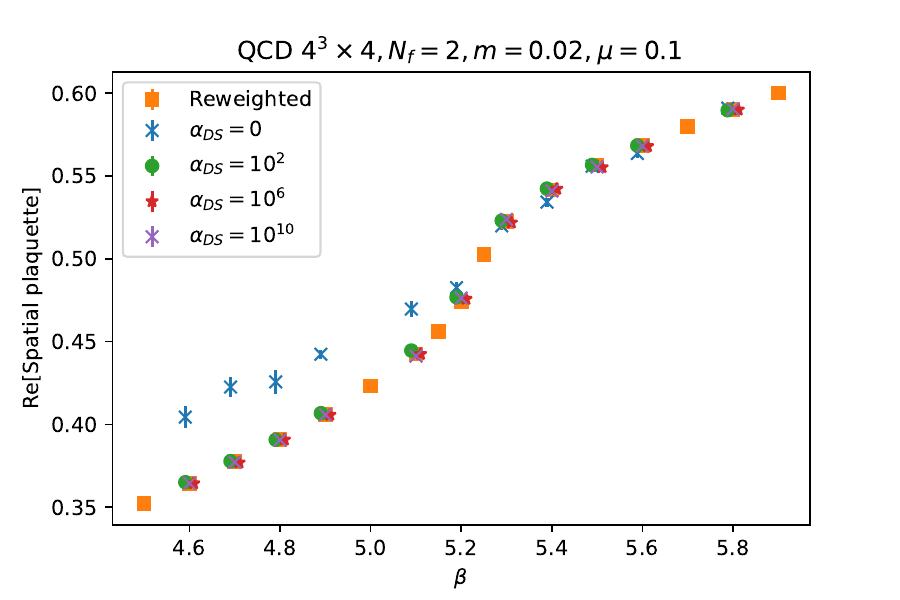}
    \includegraphics[width=0.9\columnwidth]{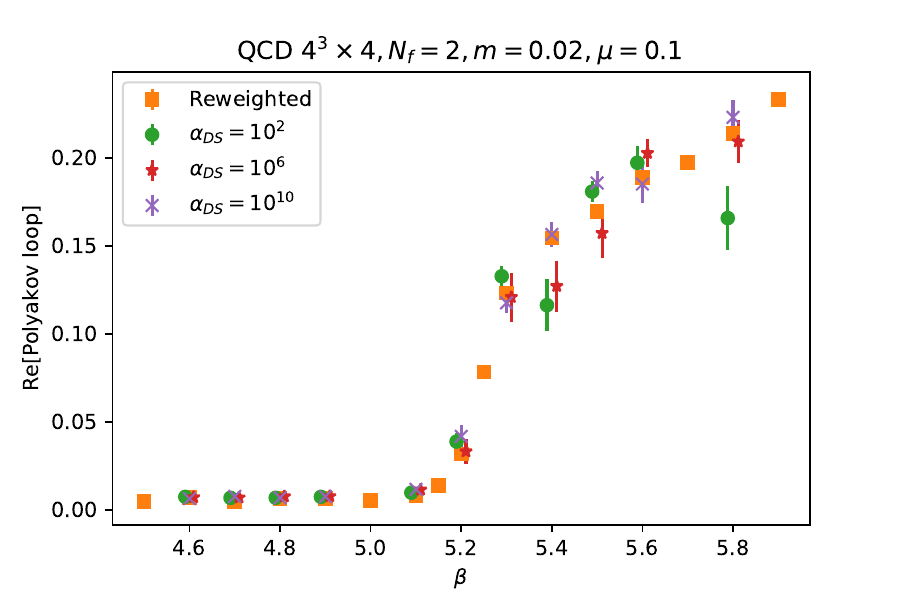}
    \includegraphics[width=0.9\columnwidth]{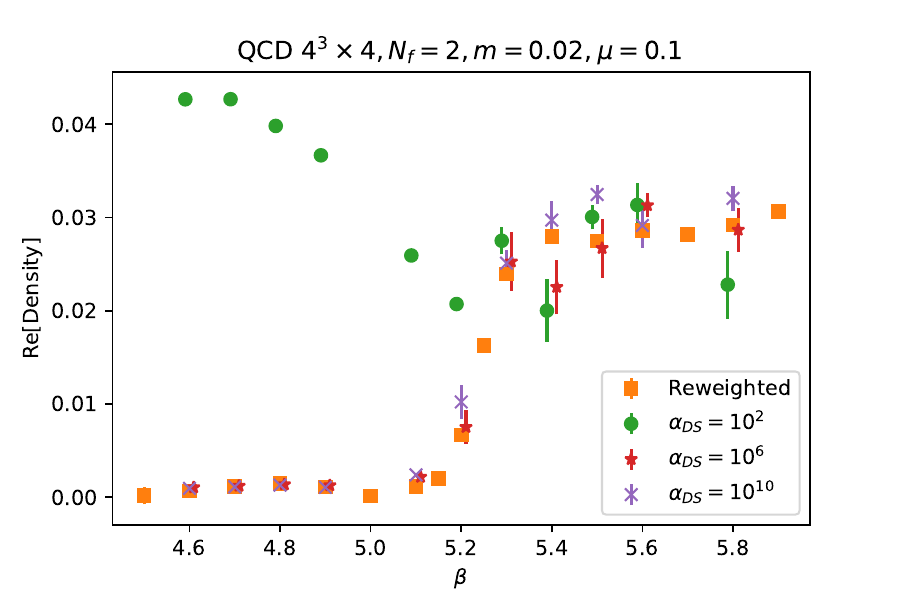}
    \includegraphics[width=0.9\columnwidth]{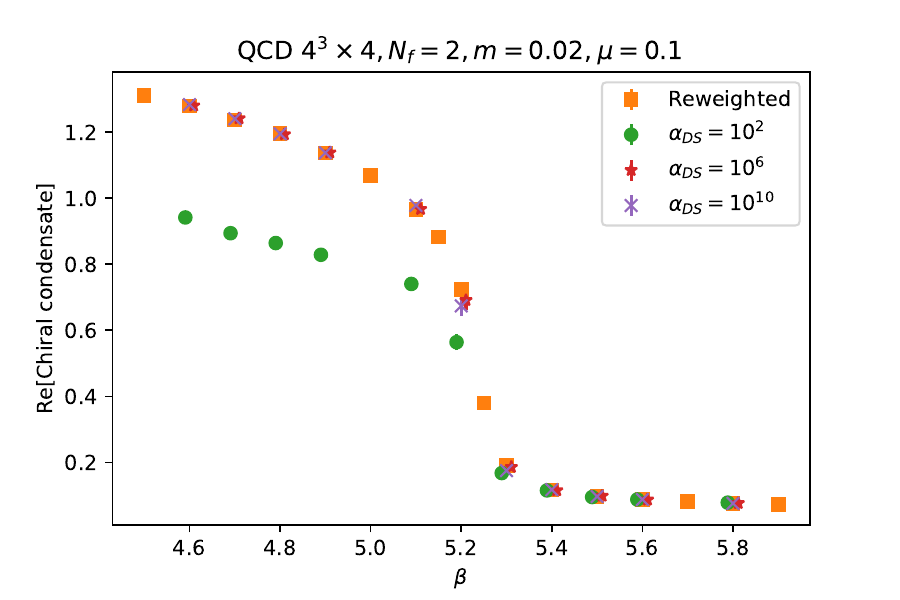}
    \caption{ Various observables (spatial plaquette, Polyakov loop, chiral condensate, baryonic density) 
    as a function of the inverse gauge coupling beta $\beta$ for QCD simulations on a $4^4$ lattice, for 
    various dynamical stabilization parameters, as indicated.
}
    \label{fig:betaplot44}
\end{figure}

We used simulations with adaptive stepsize control to get rid
of the runaway trajectories \cite{Aarts:2009dg}. 
The simulations are started from a random initial SU(3) configuration, and we 
use a $\tau=10$ Langevin time period for thermalization and 
we measure observables for a further $ \tau=90$ Langevin time. Our main 
observables are the plaquette average
\bea
 U_{\alpha\beta}={1\over N_s^3 N_t} \sum_x U_{x,\alpha} U_{x+\hat\alpha,\beta} U_{x+\hat\beta,\alpha}^{-1} U_{x,\beta}^{-1},
\eea
the Polyakov loop and its inverse
\bea
P = {1\over N_s^3 } \sum_x \Tr \left(\prod_{\tau=0}^{N_\tau-1} U_{x+\tau \hat 0,0}\right) , \\
P' = {1\over N_s^3 } \sum_x \Tr \left(\prod_{\tau=N_\tau-1}^{0} U^{-1}_{x+\tau \hat 0,0}\right),
\eea
as well as the fermionic observables: the chiral condensate 
\bea 
\chi=  {1\over \Omega}  { \partial \ln Z \over \partial m}   =  {N_F \over 4 \Omega} \Tr ( M^{-1} )
\eea 
and the baryonic density 
\bea
n_B = {1\over \Omega} { \partial \ln Z \over \partial \mu} = { N_F \over 4 \Omega } 
\Tr ( M^{-1} \partial_\mu M ).
\eea
We also measure the ensemble average of the maximal unitarity norm on the lattice.
\bea 
N_U=  {1\over 4 \Omega} \max_{x,\nu} \Tr \left[ ( U_{x\nu } U_{x \nu } ^{\dagger} -1  )^2 \right] ,
\eea
with the dimensionless space-time volume $ \Omega= N_s^3 N_t $.
$ N_U \ge 0 $ is not a physical observable, it  monitors how far the link variables
are from the SU(3) manifold (we have $N_U=0$ for an SU(3) configuration).

To get a broad overview of the effect of dynamical stabilization
we first performed simulations on $4^4$ lattices.
For these simulations we used an exact drift force for 
the fermionic drift terms 
(instead of the noisy estimator described in \cite{Sexty:2013ica}).
The exact drift terms allow for simulations at low $\beta$ values 
also at small $\ads$ values, but the numerical costs are considerably 
higher \cite{upcoming}, therefore we restricted the lattice size to $4^4$. (The noisy estimator at low $\beta$
without substantial stabilization leads to instable simulations.)
In Fig.~\ref{fig:betaplot44} we show the plaquette average 
as well as the Polyakov loop, chiral condensate and the baryonic
density. We observe that no stabilization or not 
enough stabilization at low $\beta$ values gives 
incorrect results for all observables, while at high $\beta$ (corresponding
to high temperatures) we see correct results. Introducing the stabilization,
we get seemingly correct results also at low $\beta$.
Next, we precisely calculate 
the observables on a larger lattice ($8^3 \times 4 $) to see if there
is some small bias remaining, caused by the non-holomorphic 
stabilization force.

\begin{figure}
    \centering
    \includegraphics[width=0.9\columnwidth]{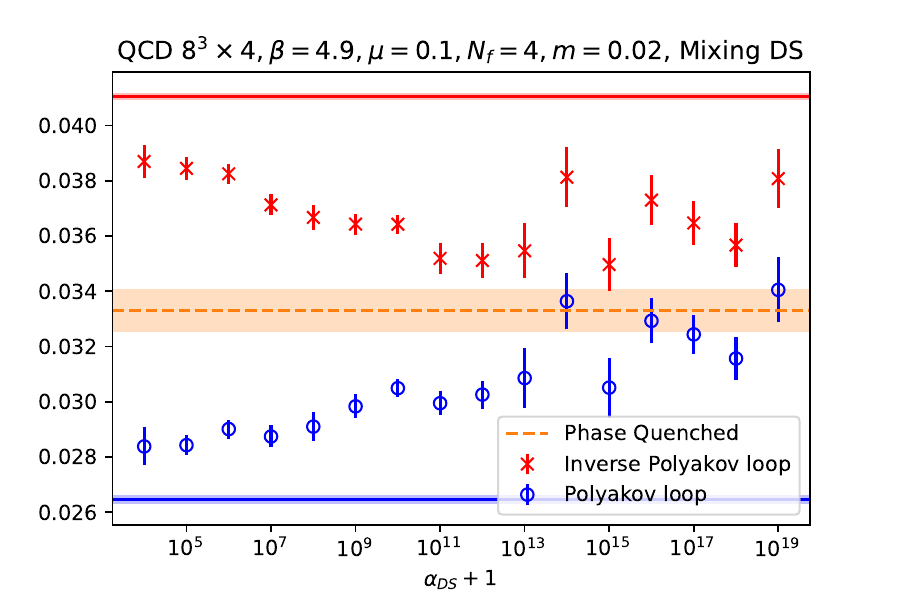}
    \includegraphics[width=0.9\columnwidth]{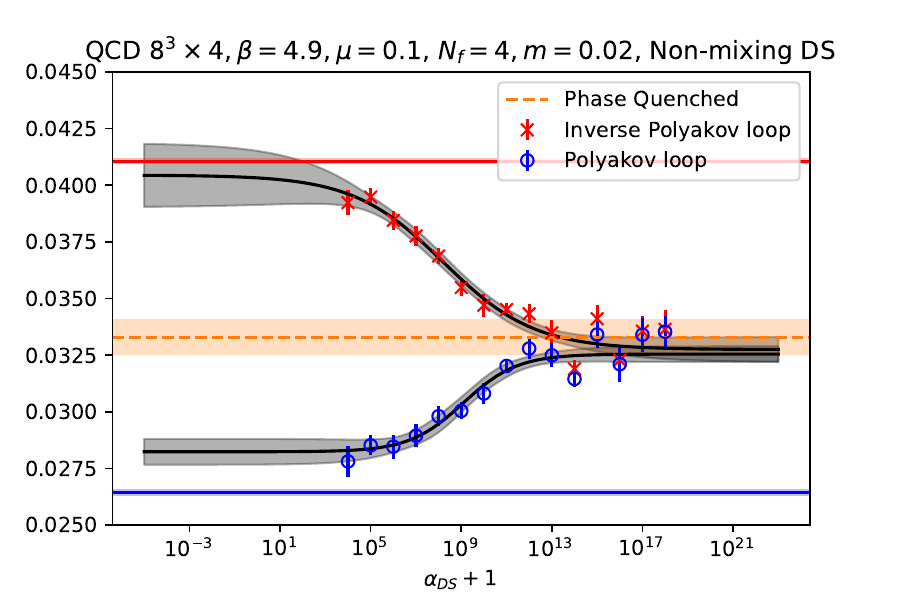}
    \caption{The Polyakov Loop $P$ and its inverse $P'$ in QCD simulations in
    the low temperature, confined phase of the theory at $\mu=0.1$.
    Lines represent results reweighted from $\mu=0$. Also
    shown is the phase quenched result. The CLE simulations use 
    the mixing formulation (\ref{mixingdynstab}) of the dynamical stabilization force above, and the non-mixing (\ref{nonmixingdynstab}) below.}
    \label{fig:rewcomplowt}
\end{figure}

In Fig.~\ref{fig:rewcomplowt} we show results from the low temperature phase 
of the theory. In this phase the CLE simulations at $\ads=0$ (or too small $\ads$) 
tend to be 
unstable, introducing a small $\ads$ allows for long simulations.

\begin{figure}
    \centering
    \includegraphics[width=0.9\columnwidth]{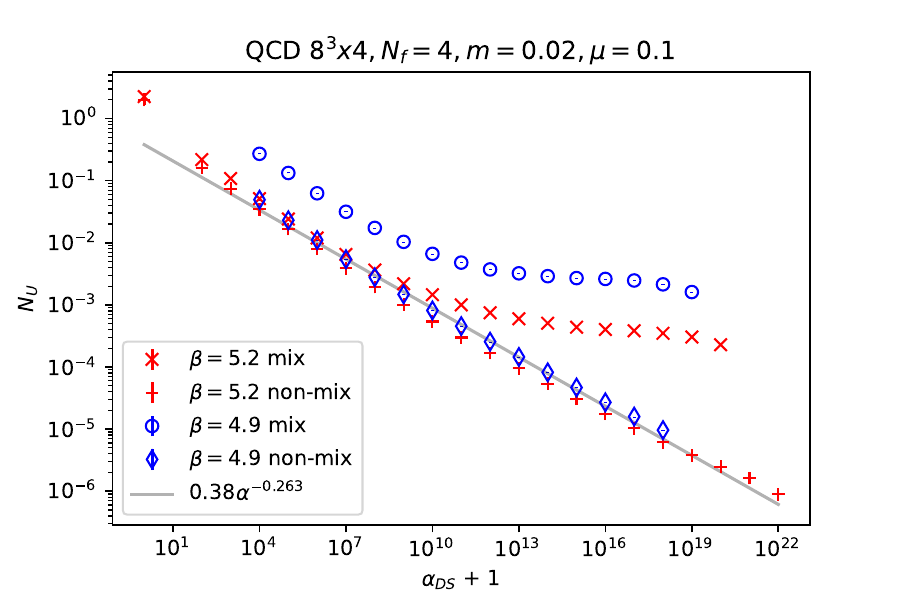}  
       \includegraphics[width=0.9\columnwidth]{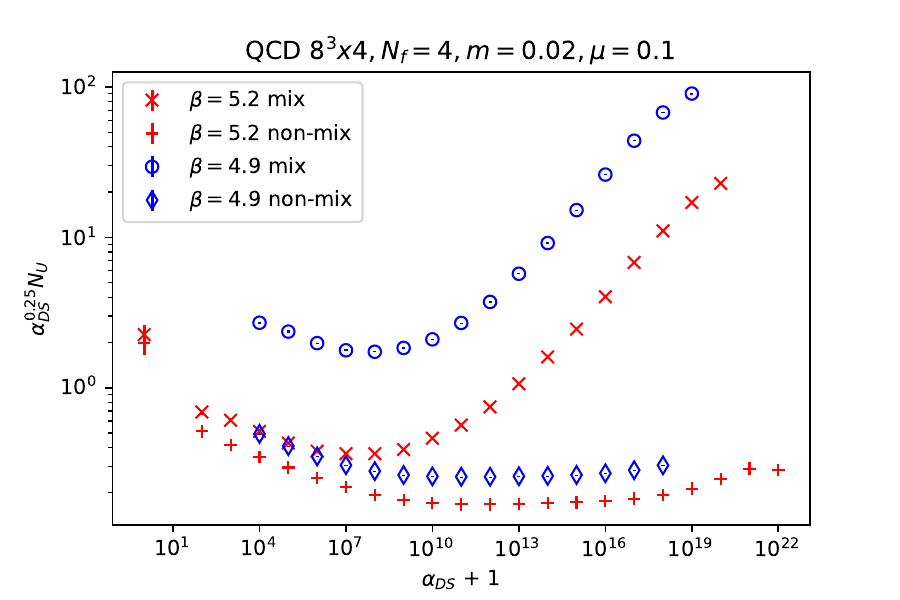}   
    \caption{The unitarity norm as a function of $\ads$ in QCD simulations at low temperatures (blue symbols) and
    high temperatures (red symbols). Two versions of the dynamical stabilization force are used, as indicated.
      Below, the average unitarity norm rescaled with $ \ads^{0.25} $, 
    as indicated. 
    }
    \label{fig:un_norms}
\end{figure}
\begin{figure}
    \centering
    \includegraphics[width=0.9\columnwidth]{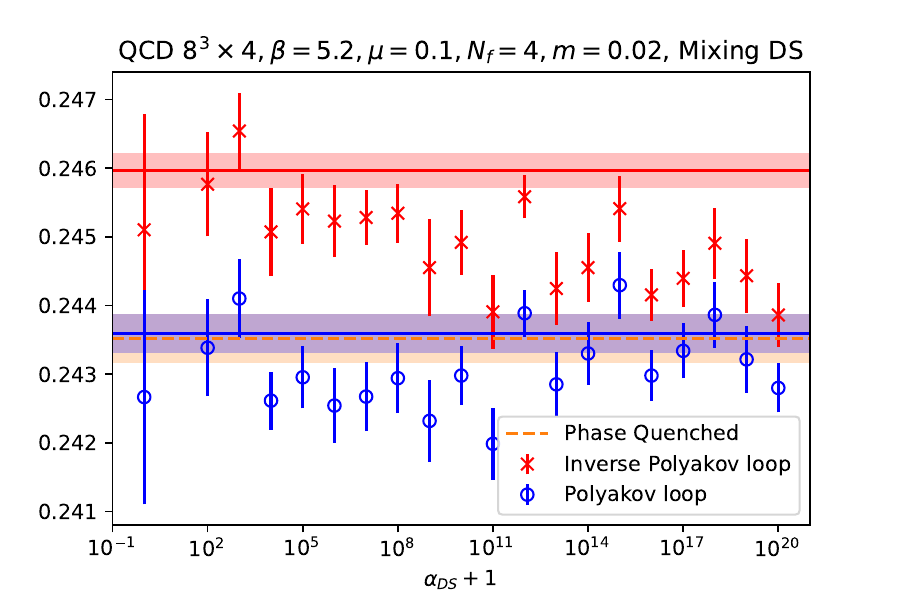}    
    \includegraphics[width=0.9\columnwidth]{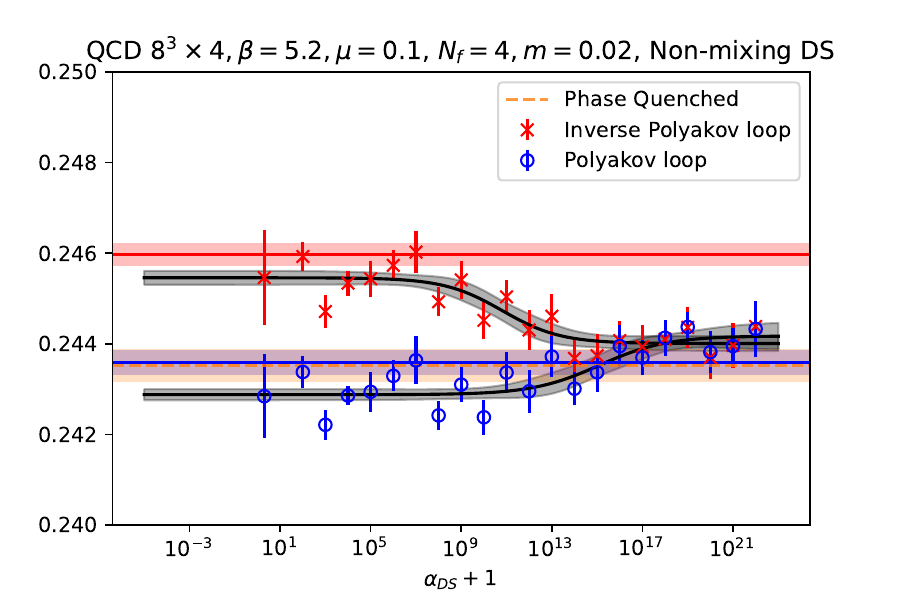}
    \caption{
    The Polyakov Loop $P$ and its inverse $P'$ in QCD simulations in
    the high temperature, deconfined phase of the theory, at $\mu=0.1$.
    Lines represent results reweighted from $\mu=0$. Also
    shown is the phase quenched result. The CLE simulations use 
    the mixing formulation (\ref{mixingdynstab}) of the dynamical stabilization force above, and the non-mixing (\ref{nonmixingdynstab}) below.
    }
    \label{fig:rewcomphight}
\end{figure}

One observes that for moderate $\ads$ the results are close to the 
exact results for both versions of the stabilizing force, but in the case of the inverse Polyakov loop we see 
statistically significant differences. At large $\ads$
the results get nearer to the phase quenched values, as expected.
We note that the non-mixing dynamical stabilization seems to present 
a stronger push towards the real manifold, as we observe that for 
larger $\ads$ the Polyakov loop and the inverse Polyakov loop
are approximately equal, signalling that the link variables
are close to the SU(3) manifold, such that effectively we 
simulate the phase-quenched system.
This is confirmed by Fig.~\ref{fig:un_norms},
where the average unitarity norm is shown as a function of $\ads$.
We note that at very large $\ads$ the simulation times 
increase as the dynamical stabilization force gets higher and in turn the adaptively
controlled Langevin stepsize can get very small. 
In Fig.~\ref{fig:rewcomphight} we show the comparison of the Polyakov loops with reweighting
results for high temperature, above the deconfinement transition for $ \mu=0.1$.
In this phase the simulation already at $\ads=0$ supplies results consistent with the exact results, i.e. dynamical stabilization is not needed. If one uses a too high 
$\ads$, the results get close to the phasequenched results, as expected.
\begin{figure}
    \centering
    \includegraphics[width=0.9\columnwidth]{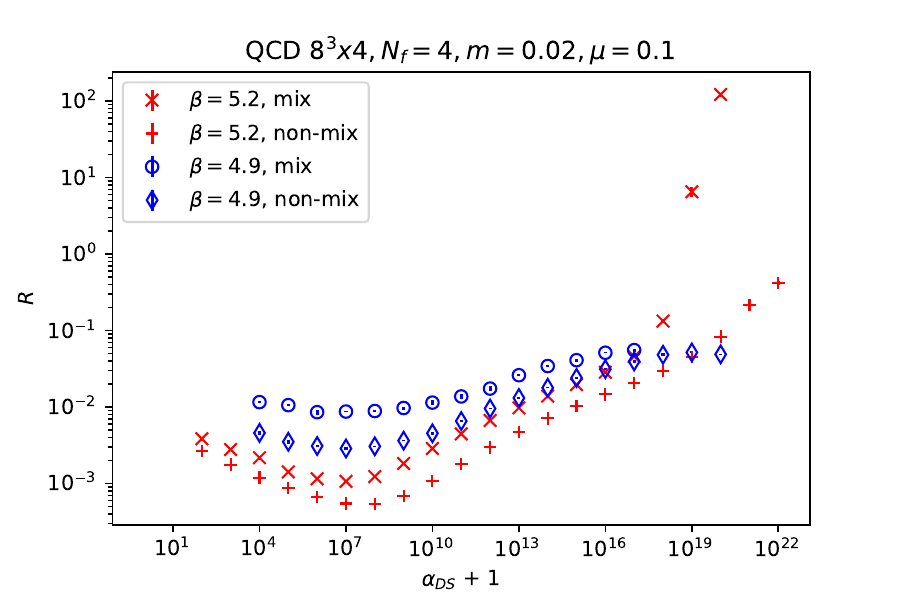}    
    \caption{The ratio of the norm for the drift terms coming from the action
    and the stabilizing force 
    (defined in (\ref{Rdefinition})) 
    in the QCD simulations is shown
    as a function of $\ads$ in QCD simulations at different temperatures
    and two versions of the dynamical stabilization force are used, as indicated.}
    \label{fig:R-QCD}
\end{figure}
We fit the results using the sigmoid ansatz (\ref{fitfunction}), 
and we get relatively good quality fits with $ \chi^2/n_\textrm{DOF} \approx 1-2.3$.
In Fig.~\ref{fig:R-QCD} we show the ratio of the norms of the "normal" drift terms 
comeing from the action and the dynamical
stabilization force, as defined in (\ref{Rdefinition}).
The norm of the "normal" part of the drift is roughly independent of $\ads$, 
so this ratio 
essentially measures how large the stabilization forces are. First of all,
we see that the ratio is always relatively small, meaning that the dynamical  stabilization
is indeed a small correction to the drift.
This is especially true for high temperatures, where simulations without 
stabilzation can work as well. For lower temperatures the 
stabilizing force gets relatively larger, but they are at least an order of magnitude smaller than
the force comeing from the action (for not too large $\ads$ values). One also observes generally larger stabilization terms 
for the mixing stabilization, this is a consequence of the slightly more 
spread out distribution for the mixing stabilization for a given
$\ads$, as observed in Fig.~\ref{fig:un_norms}.


Finally we note that for the non-mixing stabilization force the decay of the average unitarity norm
as $\ads$ is increased is approximately described by a power law dependence $ N_U \sim \ads^{-0.25}$. For the mixing version
this dependence is more complicated, the rate of decay gets diminished for large $\ads$,
as visible in Fig.~\ref{fig:un_norms}.

\begin{figure}
    \centering
    \includegraphics[width=0.9\columnwidth]{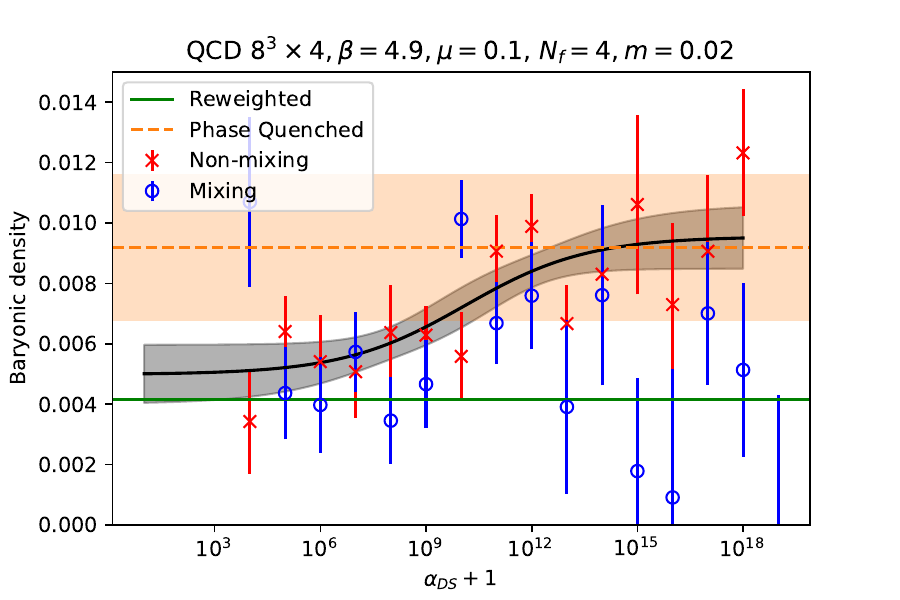}  
    \includegraphics[width=0.9\columnwidth]{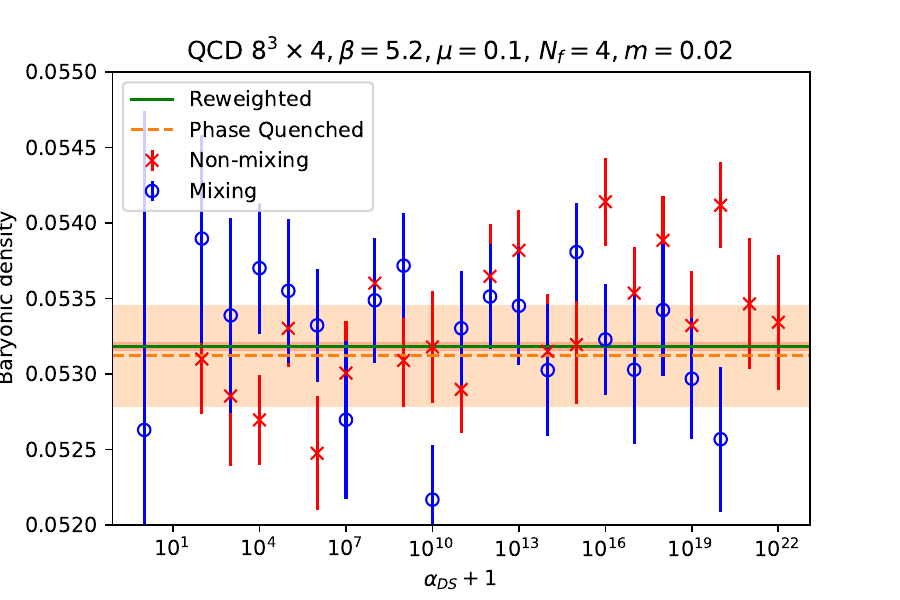}
    \caption{The baryonic density in QCD simulations at low temperatures (above) and high temperatures (below) as 
    a function of $\ads$. The low temperature CLE results with non-mixing stabilization (red symbols) are fitted with 
    a sigmoid curve (\ref{fitfunction}), where we have kept the $D$ parameter at a fixed value $D=0.25$. }
    \label{fig:dens}
\end{figure}

In Fig.~\ref{fig:dens} we show the fermionic density. At low temperatures we can see a small effect of the phasequenching, and the 
simulation values follow the sigmoid fit for the non-mixing version of the stabilization force. (For the fit we had to fix the parameter
$D$ to the value $D=0.25$, to let the fit converge in spite of the biasing effect being in the same order of magnitude 
as the errorbars of the observable.) At high temperatures at the chemical potential we used here there is no significant difference 
between the phasequenched and the reweighted density. Similarly, when investigating the chiral condensate one observes 
that the CLE simulations, the reweighted results and the phasequenched results are consistent with each other within
the statistical precision we have.


\section{Conclusions}

In this paper we have investigated the proposed dynamical stabilization force
for the Complex Langevin simulations of QCD. We introduced a second version
of the stabilizing force which does not mix together the forces 
on the links pointing to the different directions attached to the same lattice point,
but deals with each link separately. The strength of the force pushing the fields towards the SU(3) manifold
is controlled by the parameter $\ads$.
This parameter needs to be changed over many orders of magnitude to change
the behaviour of the stabilized system qualitatively, so results are 
considered to be dependent on the logarithm of $ \ads$.

As the stabilizing force is an ad-hoc addition to the drift term in the Langevin
equation, this breaks the action principle and holomorphy as well. 
Such a change is expected to spoil the correspondence 
between the original ensemble on the real manifold and the complex 
Langevin results.
First we investigated the effects of dynamical stabilization on a one-link toy-model to 
gain insight in the qualitative behavior of the system. 
We observed that the effect of the stabilizing force can be understood as a soft cutoff,
such that increasing the control parameter of the stabilizng force does eventually confine the system
to the real SU(3) manifold, which is then equivalent to phasequenching.
We observed that the functional behavior of the results for varying $\ads$ is well described 
by a sigmoid fit, in a range where the stabilization of the system is not 
too weak. 
The fit function can then be extrapolated to zero stabilization 
which gives results consistent with the exact results of the system, 
in spite of CLE simulations at $\ads =0$ giving incorrect results for some 
parameter values of the action of the model.

Next we investigated QCD simulations with four degenerate flavors of staggered fermions.
For the simulations we chose a moderate chemical potential such that results can
be compared to reweighting. 
At high temperatures, simulations at $\ads=0$ already bring results consistent with 
the reweighted ones (as observed also in \cite{Fodor:2015doa}).
Simulations with a substantial $\ads$ drive the system towards the phasequenched system. Smaller
$\ads$ might be used to improve the stability 
of the simulations somewhat (such that larger stepsizes are also usable), and the small biasing might be removed 
with a sigmoid fit similar to the toy-model, if the non-mixing version of the stabilization is used.

At low temperatures, CLE simulations become instable and break down when the unitarity norm 
gets O(1). A nonzero $\ads$ makes stable simulations possible also in this 
regime. However, in this regime the DS force gives a small but nonzero bias in the 
results of the simulations, which increase as $\ads$ is increased. As expected, the results 
eventually are driven to the phasequenched results.
The original, mixing version of the stabilization
force appears weaker, and numerical problems prevent simulations 
at very high $\ads$ values (too small stepsizes would be needed), such that 
only a trend towards the phasequenced results is visible.
For the second, non-mixing version 
of the stabilizing force, similarly to high temperatures, 
a clean transition to the phasequenched results 
is visible and the behavior of the system 
is well described with a sigmoid fit.  
Extrapolating to zero stabilization 
removes most of the bias with respect to 
the exact results.

In summary, we have shown that the dynamical stabilization is a very useful tool 
for complex Langevin simulations. For a toy model its effect on the system  
is well understood and extrapolation to zero stabilization gives correct results.
We also studied application of the method to QCD, where we introduced a slightly 
modified stabilization force, the behavior of which is very similar to the 
toy-model results, such that extrapolation to zero stabilization is feasible.

\begin{acknowledgments}
We thank Michael Mandl for his comments on the manuscript.
D.~S.~ acknowledges the support of the Austrian Science Fund (FWF) through the Stand alone Project P36875.
The computational results presented have been achieved in part using the 
Vienna Scientific Cluster (VSC), and in part on the computing cluster of the University of Graz (GSC).
\end{acknowledgments}

\bibliographystyle{apsrev}
\bibliography{cites} 





\end{document}